\documentclass{llncs}

\usepackage[T1]{fontenc}
\usepackage{graphicx}
\usepackage{booktabs}
\usepackage{amsmath,amssymb}
\usepackage{multirow}
\usepackage{algorithm}
\usepackage{algpseudocode}
\usepackage{hyperref}
\usepackage{xcolor}
\usepackage{enumitem}
\usepackage{subcaption}
\usepackage{float}

\usepackage{xspace}

\begin{document}

\title{Executable Arbitrage and Market Efficiency in Prediction Markets}

\author{Jonas Gebele \and Timm Mutzel \and Prof. Dr. Florian Matthes}
\authorrunning{J. Gebele et al.}

\institute{Technical University of Munich, Munich, Germany\\
\email{\{jonas.gebele,timm.mutzel,matthes\}@tum.de}}

\maketitle

\begin{abstract}
Deterministic payoff identities imply no-arbitrage bounds in winner-takes-all
prediction markets, but violations of these bounds need not be exploitable
before settlement. We distinguish \emph{payoff-space no-arbitrage}, which
follows from terminal payoffs, from \emph{protocol-executable no-arbitrage},
which depends on the position transformations available to traders.
Polymarket's negative-risk markets make this distinction observable: linked
binary markets represent mutually exclusive outcomes, while the NegRisk
Adapter operationalizes only the NO-to-YES direction before settlement.
We reconstruct depth-aware executable portfolio values and combine them with
actor-level transaction histories and on-chain conversion traces to measure
payoff-bound violations and exploitation. Our reconstruction estimates \$1.12
million in arbitrage profit across two realization channels: \$1.086 million
from converter-enabled strategies and \$32 thousand from settlement-based
basket formation. In the CLOB sample, positive violations concentrate on the
unsupported YES side, whereas adapter-supported NO-side violations are
substantially less frequent and shorter-lived. These patterns are consistent
with the view that pre-settlement conversion strengthens enforcement by
reducing capital lock-up and enabling inventory recycling.
Finally, we implement a prototype bidirectional extension of the NegRisk
Adapter that makes the reverse path executable before settlement.
Together, our findings show that market efficiency depends not only on payoff
structure, but also on whether protocols expose payoff equivalences as
executable primitives.
\end{abstract}

\section{Introduction}

Prediction markets turn dispersed beliefs into tradable state-contingent claims whose prices are often interpreted as probabilities \cite{wolfersZitzewitz2004,wolfersZitzewitz2006}. Prior work therefore studies whether these prices forecast outcomes, are well calibrated, or aggregate information efficiently, including in recent tokenized markets where market microstructure and settlement frictions shape prices \cite{tsangYang2026,dubach2026,settlementDiscount2026}. We study a different notion of efficiency: whether prices within winner-takes-all outcome sets satisfy the no-arbitrage constraints implied by their deterministic payoff structure.

In a winner-takes-all event with $n$ mutually exclusive outcomes, a complete YES basket pays one unit of collateral at settlement, while a complete NO basket pays $n-1$ units. These payoff identities imply simple no-arbitrage bounds, but do not necessarily define executable arbitrage opportunities. Some violations can be realized immediately through protocol-supported position transformations, whereas others require traders to assemble complete baskets and hold them until settlement. Terminal payoff equivalence is therefore distinct from \emph{pre-settlement enforceability}.

Polymarket negative-risk markets provide a natural setting in which this distinction is directly observable. They link multiple binary markets into a mutually exclusive outcome set, creating event-level payoff identities across otherwise separate central limit order books (CLOBs). Polymarket's NegRisk Adapter operationalizes one direction of this payoff equivalence before settlement by converting eligible NO-side portfolios into collateral and complementary YES exposure. The reverse YES-side equivalence has no analogous conversion path and is therefore typically enforceable only through complete-basket formation and settlement.

This asymmetric realization mechanism provides a within-protocol comparison.
\emph{Settlement-based arbitrage} requires assembling an underpriced complete
basket and waiting for event resolution. \emph{Converter-enabled arbitrage}
instead uses a protocol-supported transformation to realize payoff-equivalent
outputs before settlement, allowing collateral to be recycled immediately. We
ask whether this asymmetry in realizability is reflected in the incidence,
persistence, and exploitation of payoff-bound violations.

To answer this question, we reconstruct depth-aware executable portfolio values
from historical automated-market-maker states and event-time CLOB order books,
and combine them with actor-level transaction histories and on-chain conversion
traces. In the CLOB sample, positive violations concentrate on the unsupported
YES side, whereas adapter-supported NO-side violations are substantially less
frequent. The limited duration evidence is also directionally consistent with
faster closure on the supported side. At the actor level, we estimate
\$1.12 million in arbitrage profit across the two realization channels:
\$1.086 million from converter-enabled strategies and \$32 thousand from
settlement-based complete baskets. Together, these findings are consistent with
the NegRisk Adapter changing not the underlying payoff identity, but how quickly
and capital-efficiently traders can enforce it. The paper makes three contributions:
\begin{enumerate}[leftmargin=*,itemsep=0.1ex,topsep=0.3ex]
    \item \textbf{Conceptually}, we distinguish \emph{payoff-space
    no-arbitrage} from \emph{protocol-executable no-arbitrage} and show how
    realization mechanisms determine which payoff relations are directly
    enforceable before settlement.

    \item \textbf{Empirically}, we combine depth-aware market-state measurement
    with actor-level transaction reconstruction and on-chain conversion traces
    to measure payoff-bound violations, their persistence, and their
    exploitation through converter-enabled and settlement-based strategies.

    \item \textbf{For protocol design}, we design, implement, and benchmark a
    bidirectional extension of Polymarket's NegRisk Adapter, demonstrating that
    the reverse YES-to-NO path can be made executable before settlement.
\end{enumerate}

\section{Background}

\subsection{Tokenized Binary Claims}

Polymarket represents binary markets using collateral-backed YES and NO tokens.
A unit YES token pays one unit of collateral if the corresponding outcome occurs,
while a unit NO token pays one unit otherwise
\cite{polymarketDocs,gnosisCTF}. Let $Y_i$ and $N_i$ denote the unit terminal
payoffs in market $i$. Collateral backing implies the within-market identity
\[
    Y_i + N_i = 1.
\]
The Conditional Tokens Framework (CTF) implements this identity through three
basic operations. One unit of collateral can be split into complementary YES
and NO tokens, a complete pair can be merged back into collateral, and winning
tokens can be redeemed after resolution \cite{polymarketDocs,gnosisCTF}.
These operations govern position creation, recombination, and settlement within
a single binary market.

Winner-takes-all events add structure across binary markets. They link outcomes
that are mutually exclusive and exhaustive, such that exactly one YES claim
pays one unit of collateral. This creates cross-market payoff restrictions that
do not follow from the local identity $Y_i+N_i=1$ alone. In particular,
exclusivity constrains the joint value of claims across the complete outcome
set. Section~3 formalizes these event-level identities.

\subsection{Market Architecture}

Polymarket initially used fixed-product market maker (FPMM) contracts based on
constant-product pricing before transitioning to a central limit order book
(CLOB) architecture in late 2022
\cite{umaCLOBFunding,polymarketDocs}. In the FPMM regime, executable trade
values depend on pool reserves and trade size. In the CLOB regime, they depend
on resting limit orders and available order-book depth.

The current architecture separates order management from settlement. The order
book and matching engine operate off-chain, while token transfers, collateral
settlement, and CTF operations are executed on-chain through exchange smart
contracts \cite{polymarketDocs}. This provides two complementary data sources:
off-chain order books determine the liquidity available at executable prices,
while on-chain transactions record realized position and collateral movements.

\begin{figure}[t]
    \centering
    \begin{minipage}[t]{0.48\textwidth}
        \centering
        \vspace{0pt}
        \includegraphics[width=\textwidth]{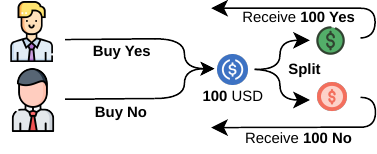}
    \end{minipage}
    \hfill
    \begin{minipage}[t]{0.48\textwidth}
        \centering
        \vspace{0pt}
        \includegraphics[width=\textwidth]{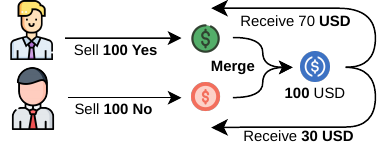}
    \end{minipage}
    \caption{Split- and merge-assisted matching in Polymarket's unified order book. Complementary buy orders are settled through a CTF split (left), while complementary sell orders are settled through a CTF merge (right).}
    \label{fig:ctf-split-merge}
\end{figure}

Polymarket presents complementary YES and NO views of an economically unified
binary order book. Buying YES at price $p$ is economically equivalent to selling
NO at price $1-p$, allowing orders across both views to be matched. As shown in
Figure~\ref{fig:ctf-split-merge}, complementary buy orders can be settled by
splitting collateral, whereas complementary sell orders can be settled by
merging complete positions.

Raw \texttt{OrderFilled} events do not by themselves provide a normalized
representation of economic exposure in split- and merge-assisted matches.
We therefore interpret fills jointly with the CTF operations executed in the
same transaction, as detailed in Section~\ref{sec:methodology}.

\subsection{Negative-Risk Conversion}

Negative-risk events connect these binary markets through event-level
exclusivity. A NO position on one outcome is payoff-equivalent to YES exposure
to the remaining outcomes, while larger NO-side subsets are equivalent to
collateral plus YES exposure on the complementary outcomes. In the earlier FPMM
regime, these relations existed as terminal payoff identities but were not
exposed through a dedicated conversion operation.

In the current CLOB regime, Polymarket's NegRisk Adapter exposes one direction
of this payoff structure as a pre-settlement token operation
\cite{polymarketDocs,negRiskAdapter}. Eligible NO-side portfolios can be
converted into collateral and complementary YES exposure before oracle
resolution. The Adapter therefore changes the realization path of an existing
payoff equivalence rather than the underlying payoff relation itself.

The conversion mechanism is asymmetric. NO-side equivalences can be realized
through the Adapter before settlement, whereas the reverse YES-to-NO direction
has no comparable protocol operation and remains settlement-dependent. The
payoff relation is therefore symmetric, but its protocol realization is not.
The next section formalizes these identities and the arbitrage opportunities
they imply.

\section{From Payoff Identities to Executable Arbitrage}

We derive the event-level payoff identities induced by mutually exclusive
outcome sets and distinguish two realization mechanisms: settlement-based and
converter-enabled arbitrage. We then derive the empirical implications of
Polymarket's one-sided conversion mechanism for the enforcement of otherwise
symmetric payoff relations.

\subsection{Deriving Event-Level Payoff Identities}

We begin with a simple three-outcome negative-risk event with mutually exclusive
outcomes $A$, $B$, and $C$. The portfolio $N_A+N_B$ pays $(1,1,2)$ across
states $(A,B,C)$, which is the same state-wise payoff as $\mathbf{1}+Y_C$.
Table~\ref{tab:three-outcome-equivalence} makes this equivalence explicit.
\begin{table}[t]
\centering
\begin{tabular}{c|c@{\hspace{1.2em}}c|c|c@{\hspace{1.2em}}c}
\toprule
\textbf{Realized outcome}
& $\boldsymbol{N_A}$
& $\boldsymbol{N_B}$
& $\boldsymbol{N_A+N_B}$
& $\boldsymbol{Y_C}$
& $\boldsymbol{\mathbf{1}+Y_C}$ \\
\midrule
$A$ & $0$ & $1$ & $1$ & $0$ & $1$ \\
$B$ & $1$ & $0$ & $1$ & $0$ & $1$ \\
$C$ & $1$ & $1$ & $2$ & $1$ & $2$ \\
\bottomrule
\end{tabular}
\caption{State-wise payoff equivalence in a three-outcome negative-risk event.}
\label{tab:three-outcome-equivalence}
\end{table}
The three-outcome example generalizes to any mutually exclusive and exhaustive
outcome set. Let $Q$ denote such a set. For each $i\in Q$, let $Y_i$ denote the
YES claim, $N_i$ the corresponding NO claim, and $\mathbf{1}$ one unit of
collateral. Throughout, $\equiv$ denotes state-wise payoff equivalence rather
than equality of market prices. The within-market complementarity relation is
\begin{equation*}
    N_i \equiv \mathbf{1}-Y_i,
\end{equation*}
and event-level exclusivity gives
\begin{equation*}
    \sum_{i\in Q} Y_i \equiv \mathbf{1}.
\end{equation*}

Combining these two relations gives the subset identity underlying
negative-risk conversion. For any subset $S\subseteq Q$, the NO basket over $S$
is payoff-equivalent to collateral plus YES exposure on the complementary
outcomes:
\begin{equation}
\label{eq:negative-risk-subset-identity}
\begin{aligned}
    \sum_{k\in S} N_k
    &\equiv \sum_{k\in S}\left(\mathbf{1}-Y_k\right) \\
    &\equiv |S|\mathbf{1}-\sum_{k\in S}Y_k \\
    &\equiv (|S|-1)\mathbf{1}+\sum_{j\in Q\setminus S}Y_j .
\end{aligned}
\end{equation}
This identity is a statement about terminal payoffs. Whether it becomes an
arbitrage opportunity depends on executable prices and on whether the protocol
provides a pre-settlement realization path.

\subsection{Settlement-Based and Converter-Enabled Realization}

Settlement provides the baseline realization path for complete baskets. The two
relevant complete-basket identities are
\begin{equation*}
    \sum_{i\in Q}Y_i \equiv \mathbf{1},
    \qquad
    \sum_{i\in Q}N_i \equiv (|Q|-1)\mathbf{1}.
\end{equation*}
If a complete YES basket costs less than one unit of collateral, or a complete
NO basket costs less than $|Q|-1$ units, a trader can acquire the basket, hold it
until oracle resolution, and redeem its deterministic payoff.

Let $a_t(P)$ denote the executable acquisition cost of a unit-sized portfolio
$P$ at time $t$. In the simplest unit-basket representation, the corresponding
settlement edges are
\begin{equation}
\label{eq:settlement-edges}
    \Delta_Y^{\mathrm{settle}}(t)
    =1-\sum_{i\in Q}a_t(Y_i),
    \qquad
    \Delta_N^{\mathrm{settle}}(t)
    =(|Q|-1)-\sum_{i\in Q}a_t(N_i).
\end{equation}
A positive edge identifies an underpriced complete basket. Realizing this edge
through settlement locks the invested capital until the oracle finalizes the
event.

The NegRisk Adapter provides an additional pre-settlement realization path for
NO-side portfolios. For any subset $S\subseteq Q$, it operationalizes
Equation~\eqref{eq:negative-risk-subset-identity} in the direction
\[
    \sum_{k\in S}N_k
    \longrightarrow
    (|S|-1)\mathbf{1}
    +\sum_{j\in Q\setminus S}Y_j.
\]
A trader can therefore acquire the NO-side input portfolio, convert it, and
immediately recover collateral and complementary YES exposure. Let $b_t(P)$
denote the executable sale proceeds from portfolio $P$. The corresponding
unit-size edge is
\begin{equation}
\label{eq:no-to-yes-edge}
    \Delta_S^{N\rightarrow Y}(t)
    =
    (|S|-1)
    +\sum_{j\in Q\setminus S}b_t(Y_j)
    -\sum_{k\in S}a_t(N_k).
\end{equation}

The same payoff identity can be priced in the reverse direction:
\begin{equation}
\label{eq:yes-to-no-edge}
    \Delta_S^{Y\rightarrow N}(t)
    =
    \sum_{k\in S}b_t(N_k)
    -\sum_{j\in Q\setminus S}a_t(Y_j)
    -( |S|-1 ).
\end{equation}
A positive reverse edge would be realizable before settlement if a trader could
supply the complementary YES portfolio and required collateral in exchange for
the NO portfolio over $S$. Polymarket currently provides no such transformation.

The empirical analysis generalizes these expressions to depth-aware,
fee-adjusted portfolio values. Settlement-based realization requires complete
baskets and delayed redemption, whereas the existing conversion path realizes
NO-to-YES subset identities while the event remains open.

\subsection{Asymmetric Enforcement}

The payoff relation in
Equation~\eqref{eq:negative-risk-subset-identity} is symmetric, but its
protocol-level realization is not. Polymarket exposes the NO-to-YES direction
before settlement, whereas the reverse YES-to-NO direction remains unsupported.
Payoff-equivalent deviations may therefore differ in their incidence,
persistence, and realized exploitation.

This asymmetry yields three empirical implications:
\begin{enumerate}[label=\textbf{I\arabic*.},leftmargin=*,itemsep=0.2ex,topsep=0.3ex]
    \item The converter-supported direction should exhibit a lower incidence of
    positive payoff-bound violations.

    \item Positive violations in the converter-supported direction should be
    less persistent because collateral can be recovered and recycled before
    settlement.

    \item Realized exploitation should be concentrated in converter-enabled
    strategies, whereas unsupported opportunities should be realized primarily
    through complete-basket formation and settlement.
\end{enumerate}

The empirical analysis examines whether these directional differences are
present in market states and actor-level transaction histories.

\section{Methodology}
\label{sec:methodology}

To evaluate the empirical implications derived in Section~3, we conduct two
complementary analyses. The market-state analysis reconstructs executable
payoff-bound violations and measures their persistence, while the actor-level
analysis reconstructs mechanism-linked exploitation through NegRisk Adapter
conversions and settlement-based complete baskets. Because historical level-2
order-book data are unavailable for the transaction sample, the two analyses
rely on distinct data panels and cannot be matched at the
individual-opportunity level.

\subsection{Empirical Scope and Sample Construction}

Our empirical design combines three non-overlapping data panels. The first
covers Polymarket's historical fixed-product market maker (FPMM) regime and
contains 51 mutually exclusive and exhaustive outcome sets spanning 235 binary
markets and approximately 53,000 trades. Because Polymarket did not explicitly
identify mutually exclusive market groups through its FPMM API, we derive
candidate sets from event metadata and market wording and manually validate
them against the corresponding resolution rules. We retain only sets in which
exactly one outcome could resolve YES. Using Polygon RPC data, we reconstruct
trades and pool states to obtain executable acquisition costs and sale proceeds.

The second panel contains the historical CLOB transaction data used for the
actor-level analysis. It comprises 32,702 Polymarket events and 259 million
trades for events listed through 31 December 2025 and covered by the
cross-market payoff bounds derived in Section~3. We obtain event metadata and
on-chain \texttt{OrderFilled} traces from
\cite{polymarket_data_2026}. NegRisk Adapter conversions and CTF splits and
merges are reconstructed from Goldsky subgraphs and parsed on-chain logs
\cite{goldskyPolymarketIndexing} and cross-checked against PolygonScan
\cite{polygonscanExplorer}. Throughout the analysis, an actor denotes the
address or Polymarket proxy wallet to which fills and token positions are
attributed.

Executed transactions do not reveal the prices and quantities simultaneously
available across all component markets. The third panel therefore provides
level-2 CLOB data for the market-state analysis. Because these data are
available only from April 2026, we use a separate hour-stratified observation
period from 14 April 2026 00:00 UTC to 19 May 2026 06:00 UTC
\cite{pmxtPolymarketOrderbookArchiveV2}.\footnote{In the coming weeks, we extend this observation window by 24 observation periods.} We
treat each sampled hour as an
independent observation window, initialize the books from level-2 snapshots,
and replay all incremental updates. This produces event-time order-book states
rather than isolated snapshots. At each state, we jointly walk the available
depth across all component markets to reconstruct executable portfolio values
and measure the incidence and persistence of payoff-bound violations.

We analyze the three panels separately. Comparisons between the FPMM and CLOB
regimes are therefore descriptive rather than causal before--after estimates,
as the regimes differ in market structure, trading volume, liquidity, and
observation period.

\subsection{Payoff-Bound Violation Measurement}
\label{sec:payoff-bound-measurement}

Equation~\eqref{eq:negative-risk-subset-identity} establishes the subset payoff
identity for negative-risk events. At each reconstructed market state, we
compare executable acquisition costs and sale proceeds to measure the incidence
and fee-adjusted magnitude of deviations from the associated payoff bounds. For
an event with active outcome set $Q$ and subset $S\subseteq Q$, let
$\bar S=Q\setminus S$ and define
\begin{equation*}
N_S=\sum_{i\in S}N_i,
\qquad
Y_{\bar S}=\sum_{j\in\bar S}Y_j,
\end{equation*}
so that
\begin{equation*}
N_S\equiv (|S|-1)\mathbf{1}+Y_{\bar S}.
\end{equation*}
The boundary cases $S=Q$ and $S=\varnothing$ recover the complete NO- and
YES-basket bounds in
Equation~\eqref{eq:negative-risk-subset-identity}.

Let $a_t(P)$ denote the executable cost of acquiring portfolio $P$ in market
state $t$, and let $b_t(P)$ denote the proceeds from selling it. For CLOB
markets, we obtain these values by jointly walking the ask or bid depth across
the relevant component markets after each order-book update. For FPMM markets,
we simulate the corresponding trades against each reconstructed pool state.

Using these executable values, we measure deviations in both directions. The
fee-adjusted NO-to-YES edge is
\begin{equation*}
\Delta_t^{N\rightarrow Y}(S)
=
(|S|-1)+b_t(Y_{\bar S})-a_t(N_S)-f,
\end{equation*}
and the fee-adjusted YES-to-NO edge is
\begin{equation*}
\Delta_t^{Y\rightarrow N}(S)
=
b_t(N_S)-a_t(Y_{\bar S})-(|S|-1)-f.
\end{equation*}

A positive edge indicates that the executable sale value on one side of the
identity exceeds the acquisition cost on the other, including the required
collateral transfer and net of fees. We refer to positive
$\Delta_t^{N\rightarrow Y}$ and $\Delta_t^{Y\rightarrow N}$ as NO- and
YES-side violations, respectively, and retain the largest fee-adjusted edge
across subsets for each event, direction, and update. The term
\emph{violation} describes the observed price configuration; whether it
constitutes executable pre-settlement arbitrage additionally depends on whether
the protocol supports the corresponding realization path. Here, $f$ denotes
the aggregate fees applicable to the portfolio and direction being evaluated.
We conservatively assume taker execution for all token trades, thereby
overstating execution costs whenever maker execution is feasible and
understating the corresponding potential profitability.

\subsection{Violation Episodes and Persistence}
\label{sec:violation-persistence}

Beyond incidence and magnitude, we measure how long payoff-bound violations
persist. For each event and direction, we define a violation episode as a
maximal sequence of consecutive event-time states in which the maximum
fee-adjusted edge remains strictly positive. Because the order books are
reconstructed in event time, we assume that each state prevails until the next
relevant market update. An episode opens when the edge becomes positive and
closes at the first subsequent update at which it is non-positive. Its duration
is therefore
\begin{equation*}
    D_k=t_k^{\mathrm{close}}-t_k^{\mathrm{open}}.
\end{equation*}

For the hour-stratified CLOB panel, episodes observed during both the first and
final five minutes of a window have unknown opening and closing times. We
classify these episodes as lasting at least 50 minutes and report them
separately rather than assigning them an artificial duration. The
exact-duration analysis is restricted to episodes whose opening and closing
times are observed.

For each regime and direction, we summarize exact durations using the empirical
survival function
\begin{equation*}
\widehat{\mathcal{S}}*D(\tau)
=
\frac{1}{K}
\sum*{k=1}^{K}
\mathbf{1}{D_k>\tau},
\end{equation*}
where $K$ is the number of episodes with observed durations. Thus,
$\widehat{\mathcal{S}}_D(\tau)$ gives the proportion of these episodes whose
duration exceeds $\tau$.

\subsection{Realized Exploitation}

Market-state violations identify potential opportunities but do not establish
that any actor exploited them. We therefore require transaction-level evidence
that an actor executed a strategy realizing or locking in the corresponding
payoff equivalence. We analyze two realization channels: NegRisk Adapter
conversions and settlement-based complete baskets. Because these mechanisms
leave different transaction-level evidence, they require distinct
reconstruction procedures. Here, realized exploitation refers to observed
execution of the relevant strategy, not necessarily the immediate liquidation
of every resulting position.

\subsubsection{Converter-Enabled Arbitrage}

NegRisk Adapter calls provide direct evidence that a payoff identity was
realized before settlement. Our unit of observation is a conversion bundle, for
which we recover the actor, event, input NO positions, converted quantity,
collateral released, and returned YES positions. We decompose calls into
separate bundles when their input sets or converted quantities differ.

For each bundle, we match the converted NO positions to CLOB purchases or NO
positions minted through CTF collateral splits within the five Polygon blocks
preceding the conversion, corresponding to approximately ten seconds. Each
matched acquisition batch must cover all required input positions, although a
single conversion may draw on multiple independently valid batches. We exclude
partial baskets and allocate oversized acquisitions proportionally. Direct
purchases are valued at their fill prices, while split-minted NO positions are
assigned an opportunity value based on the estimated contemporaneous bid. We
infer executable quotes from midpoints by assuming a one-cent spread at
half-cent midpoints and a two-cent spread at integer-cent midpoints. For
midpoints below three cents or above 97 cents, we instead apply a
midpoint-to-quote adjustment of 0.1 cents.

We track returned YES positions for 150 blocks, corresponding to approximately
five minutes. Sales and redemptions are valued at realized proceeds, whereas
merged and residual positions are valued at the estimated ask at the time of
conversion using the same spread assumptions. Partial dispositions are
allocated proportionally. We define output value and estimated conversion
profit as
\begin{equation*}
V_{\mathrm{out}}
=
V_{\mathrm{collateral}}+V_{\mathrm{YES}},
\qquad
\Pi^{\mathrm{conv}}
=
V_{\mathrm{out}}-C_{\mathrm{in}},
\end{equation*}
where $V_{\mathrm{YES}}$ combines realized proceeds and imputed values for
merged and residual positions, and $C_{\mathrm{in}}$ combines observed purchase
costs with the imputed opportunity value of split-minted NO positions.
Accordingly, $\Pi^{\mathrm{conv}}$ is a conversion-centered profit estimate
that combines realized proceeds with mark-to-market valuations.

\subsubsection{Settlement-Based Basket Formation}

Settlement-based exploitation leaves no dedicated realization call and must
therefore be inferred from actor-level position histories. The unit of
observation is a complete event-level YES or NO basket formed by one actor
within one event. We consider only baskets spanning all active outcomes.

We reconstruct actor-level inventories chronologically from fills and sales,
accounting for positions created or removed through CTF splits and merges. A
basket becomes complete when the actor holds a positive quantity of the
relevant token type for every active outcome. Its matched quantity $q$ is the
minimum holding across these positions, with excess inventory in individual
legs excluded. We assign matched lots using FIFO accounting and require the
formation time
\begin{equation*}
    \tau(q)=t_{\mathrm{last}}(q)-t_{\mathrm{first}}(q)
\end{equation*}
to be at most ten minutes. Alternative time thresholds and LIFO accounting
serve as robustness checks. Matched lots are removed after extraction, allowing
the same actor to form multiple baskets within an event.

Let $C_Y(q)$ and $C_N(q)$ denote the reconstructed acquisition costs. Scaling
the complete-basket identities by $q$ gives
\begin{equation*}
\begin{aligned}
    \Pi^{Y}(q) &= q-C_Y(q), \\
    \Pi^{N}(q) &= (n-1)q-C_N(q),
\end{aligned}
\end{equation*}
for an event with $n$ active outcomes. These expressions measure the terminal
profit locked in when the basket is completed, conditional on retaining the
positions through resolution. Quantities already attributed to Adapter
conversions are excluded to avoid double counting. For both realization
channels, we omit gas costs because the relevant transactions are routed
through Polymarket's relayer.

\section{Results}

\subsection{Violation Episodes and Side Asymmetries}

\begin{figure}[t]
\centering
\begin{minipage}[t]{0.48\textwidth}
\centering
\vspace{0pt}
\includegraphics[width=\textwidth]{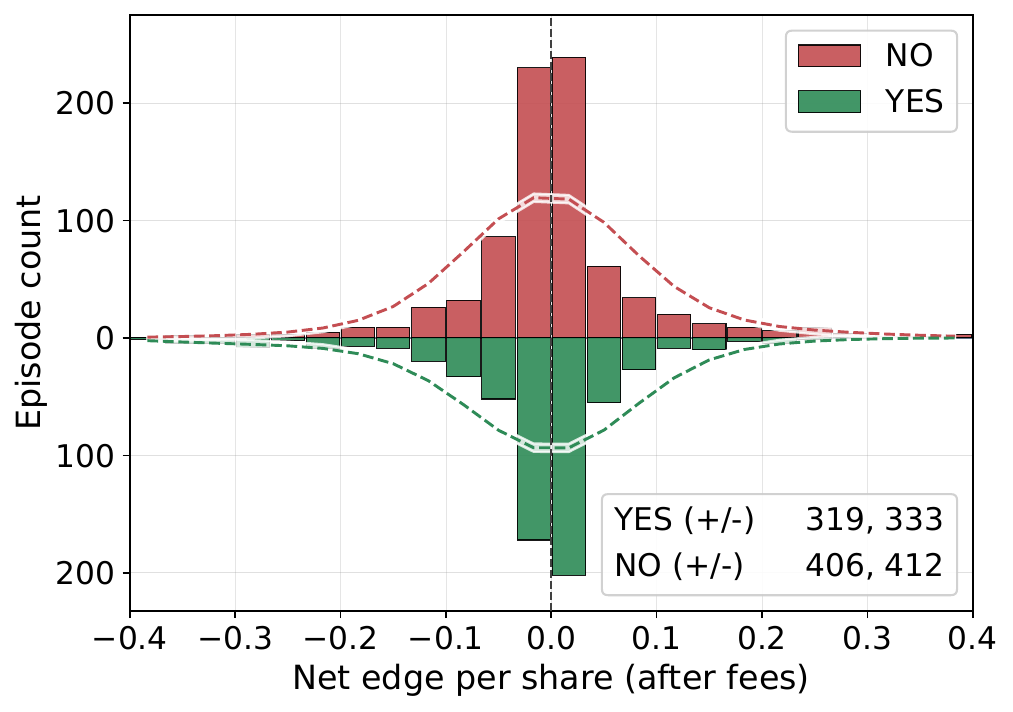}
\end{minipage}
\hfill
\begin{minipage}[t]{0.48\textwidth}
\centering
\vspace{0pt}
\includegraphics[width=\textwidth]{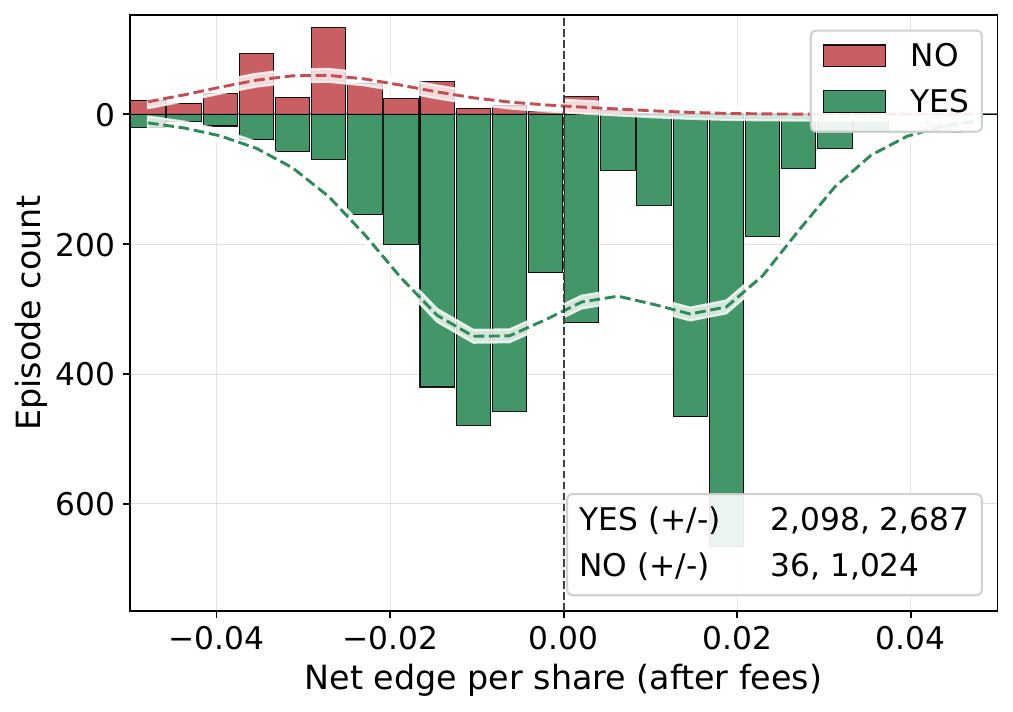}
\end{minipage}
\caption{Episode-level distributions of fee-adjusted payoff-bound edges in
the FPMM sample (left) and CLOB sample (right). Insets report the numbers of
positive and negative episodes by direction.}
\label{fig}
\end{figure}

Payoff-bound deviations differ sharply across regimes.
Figure~\ref{fig} shows broadly similar numbers
of positive and negative episodes on both the YES and NO sides in the FPMM
sample, where no explicit event-level conversion primitive was available. In
the CLOB sample, by contrast, positive episodes are overwhelmingly concentrated
on the YES side: we observe 2,098 positive YES-side episodes but only 36
positive NO-side episodes.

This concentration is consistent with the protocol's directional enforcement
asymmetry. NO-side violations can be realized through the existing NegRisk
Adapter, whereas YES-side violations would require the unavailable reverse
conversion path. Because the FPMM and CLOB panels cover different periods and
market structures, however, the comparison is descriptive and should not be
interpreted as a causal before--after estimate.

\subsection{Persistence and Closure Dynamics}

\begin{figure}[t]
\centering
\includegraphics[width=\textwidth]{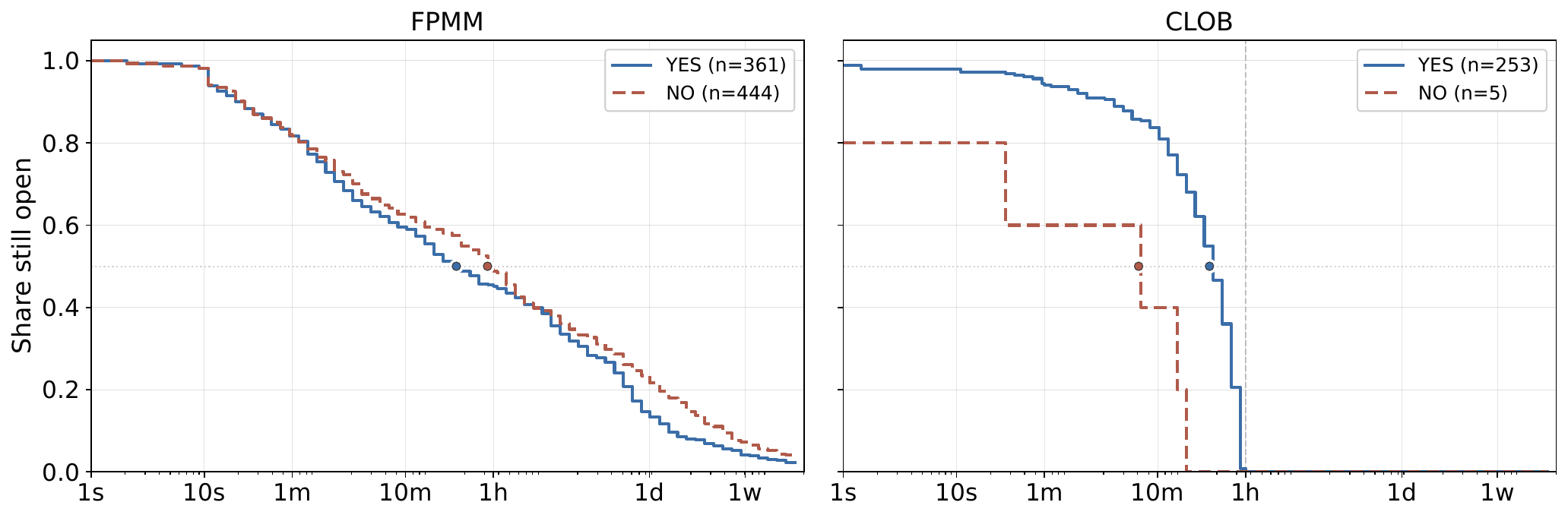}
\caption{Empirical survival functions for exact-duration, fee-adjusted
violation episodes. Dots indicate median durations; window-spanning CLOB
episodes are excluded.}
\label{fig:duration-survival}
\end{figure}

Figure~\ref{fig:duration-survival} shows substantial persistence in the FPMM
sample. The YES- and NO-side curves largely overlap, with approximately 40
episodes remaining open beyond 50 minutes and a non-negligible share persisting
for more than one day.

CLOB persistence is more difficult to assess under the hour-stratified sampling
design. Of the 624 observed episodes, 366 (58.7
first and final five minutes of their observation window. Their exact durations
are unknown, but they must exceed 50 minutes. We therefore classify them as
window-spanning and exclude them from the exact-duration survival curves.

Among the remaining 258 episodes, observed durations generally remain below
50 minutes. The median is 16.15 seconds for the 253 YES-side episodes and
7.99 seconds for the five NO-side episodes. The earlier decline of the NO-side
curve is therefore directionally consistent with converter-enabled enforcement,
but the extremely small NO-side sample does not support a reliable
distributional comparison. Moreover, episode closure indicates only that the
violation disappeared; it does not establish that an arbitrage trade caused
the closure.

\subsection{Realized Exploitation}

Across the analyzed panels, we estimate approximately 1.118 million USDC in
mechanism-linked profit. Converter-enabled arbitrage accounts for approximately
1.086 million USDC, while settlement-based basket formation accounts for
32,283 USDC. Converter-enabled realization therefore represents approximately
97

\subsubsection{Settlement-Based Arbitrage}

In the FPMM sample, aggregating the maximum fee-adjusted executable deviation
for each event yields a market-state benchmark of 5,185 USDC. The actor-level
reconstruction identifies 67 profitable complete baskets formed within ten
minutes across 19 events and 35 actors. Under FIFO cost attribution, these
baskets imply 3,639 USDC in settlement-based profit. Profit is highly
concentrated: repeat actors form 70
the profit, while one actor captures 76

In the CLOB sample, we identify 5,923 profitable complete baskets formed within
ten minutes, corresponding to 28,644 USDC in settlement-implied profit. Profit
is nearly balanced between YES baskets (14,199 USDC across 4,014 baskets) and
NO baskets (14,445 USDC across 1,909 baskets).

The NO-side estimate is notable because complete NO baskets can also be
realized before settlement through the NegRisk Adapter. The transaction traces
do not reveal why traders retained these positions rather than converting them.
Possible explanations include formation close to resolution, which reduces the
remaining capital lock-up, and batched conversion strategies that were not
executed before the event resolved. These explanations remain conjectural.
Appendix~\ref{app:settlement-basket-distribution} reports the monthly
distribution.

\subsubsection{Converter-Enabled Arbitrage}

\begin{figure}[t]
\centering
\includegraphics[width=\textwidth]{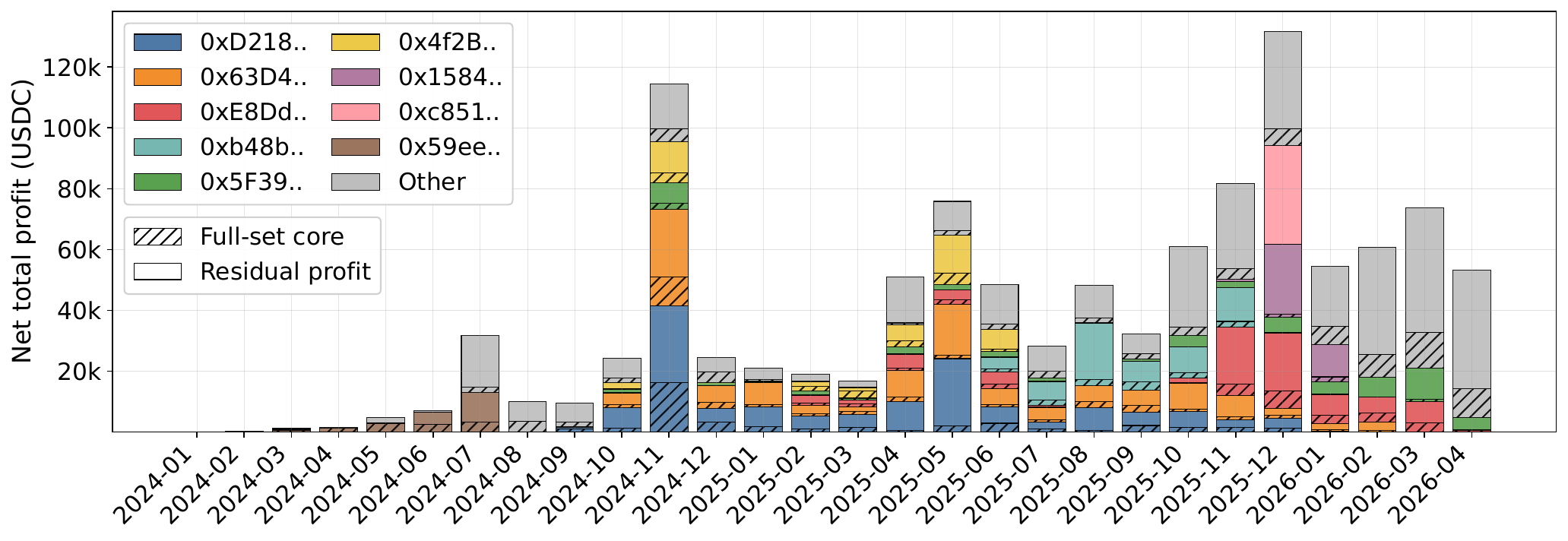}
\caption{Monthly positive net converter-enabled profit by address. Hatched
segments show core profit from full-set NO-to-collateral conversions; solid
segments show residual profit associated with partial conversions and returned
YES inventory. Only positive address-month contributions are shown.}
\label{fig:profit-per-stakeholder-over-time}
\end{figure}

Following the NegRisk Adapter's deployment on 28 November 2023,
converter-enabled profit first appears in January 2024. Our main estimates
exclude one anomalous cluster of 75 conversions executed by three addresses
between 22:07 and 22:14 UTC on 12 December 2025. The cluster produced an
estimated profit of 381,748 USDC within seven minutes, with an exceptionally
high return on input capital and nearly all profit arising from the core
conversion component. Midpoint prices remained close to 50 cents immediately
before and after the cluster, consistent with transitory orders rather than a
persistent market-wide mispricing. Taken together, these diagnostics suggest
non-representative and potentially coordinated transfers rather than ordinary
competitive arbitrage. Appendix~\ref{app:conversion-diagnostics} reports the
affected addresses, supporting diagnostics, and alternative exclusion criteria.

After this exclusion, converter-enabled profit remains concentrated among a
small set of addresses
(Figure~\ref{fig:profit-per-stakeholder-over-time}). The ten most profitable
addresses account for 75

Full-set NO-to-collateral conversions account for 205,531 USDC, or 18.9
the total 1.086 million USDC. The remaining profit is associated primarily with
partial conversions and the subsequent management or valuation of returned YES
positions. This decomposition indicates that the Adapter supports not only
direct NO-to-collateral realization but also multi-step inventory recycling.

Approximately 80
conversions are more frequent and more likely to be profitable, whereas
all-taker conversions are less frequent but generate larger profits on average.
This pattern holds for both partial and full-set conversions. Consistent with
this predominance, the weekly all-taker share in
Figure~\ref{fig:weekly-maker-taker-share} generally remains at or below 40%
from late 2024 through 2025.

The weekly profit distribution in
Figure~\ref{fig:weekly-profit-per-trade-percentiles} also changes over time.
Through July 2024, median profit is approximately one USDC per conversion, and
the interquartile range lies largely above zero. During late 2024 and 2025, the
median declines to approximately 0.20 USDC as dispersion increases and
unprofitable conversions become more common. By early 2026, median profit falls
further to approximately 0.08 USDC. Broader participation, declining typical
profits, and a persistent negative lower tail are consistent with intensifying
competition, although these descriptive trends may also reflect changes in
market composition and liquidity.

\section{Protocol Design: Bidirectional Conversion}
\label{sec:bidirectional-conversion}

The empirical results motivate a protocol-design
effect. Positive CLOB violations are substantially less frequent on the
adapter-supported NO side than on the unsupported YES side. In the limited
duration sample, NO-side episodes are also directionally shorter-lived. Moreover, reconstructed conversion
strategies show that traders use the available realization path. These findings
motivate a narrower design question: can the reverse payoff equivalence be made
executable with comparable incremental cost?

\paragraph{Reverse realization path.}
Let $Q$ denote a fixed, mutually exclusive and exhaustive outcome set, and let
$\varnothing\neq S\subseteq Q$ denote the desired NO-output set. For any
quantity $q>0$,
\begin{equation}
\label{eq:reverse-realization}
    q\sum_{j\in Q\setminus S}Y_j
    + (|S|-1)q\mathbf{1}
    \equiv
    q\sum_{i\in S}N_i .
\end{equation}
A reverse adapter operationalizes this identity from left to right: the trader
supplies the complementary YES positions and $(|S|-1)q$ collateral and receives
$q$ units of each NO position in $S$. When $|S|=1$, no collateral top-up is
required; when $S=Q$, $(|Q|-1)q$ collateral is converted into a complete NO
basket. The remaining boundary case,
$q\sum_{i\in Q}Y_i\equiv q\mathbf{1}$, converts a complete YES basket into
collateral.

\paragraph{Polymarket implementation.}
Our prototype \texttt{convertYESPositions} extends the mint-and-retire structure
of the existing NegRisk Adapter. For each $i\in S$, the Adapter synthetically
mints wrapped collateral and splits it into a YES/NO pair. The newly created YES
positions, together with the caller-supplied YES positions on $Q\setminus S$,
form a complete YES basket and are made non-redeemable. The Adapter collects the
collateral top-up and transfers the complementary NO basket to the caller.
Consequently, its net input and output satisfy
Equation~\eqref{eq:reverse-realization}. Appendix~\ref{app:algorithms} provides
pseudocode and states the associated collateral-conservation invariant.

\paragraph{Active-outcome constraint.}
Polymarket's current Adapter operates over all contract-indexed outcomes and
does not maintain a separate set of economically active outcomes. This includes
inactive placeholders. In the existing NO-to-YES direction, returning a
worthless placeholder YES position is harmless. In the reverse direction,
however, a placeholder NO position is effectively collateral-like because the
placeholder cannot win, so returning it would create unsupported value. A
reverse adapter must therefore maintain a finalized active-outcome set and
restrict conversions to that set. Our prototype establishes feasibility
conditional on this additional state-management mechanism.

\paragraph{Implementation cost.}
Across benchmark configurations with $5$--$196$ outcomes,
\texttt{convertYESPositions} consumes $63{,}095$ more gas than
\texttt{convertPositions}. Thus, the reverse path adds
nearly constant incremental overhead and preserves the linear
scaling of the existing implementation. Total gas nevertheless increases from
approximately $0.70$ million to $25.32$ million over this range, so large
outcome sets remain constrained by the absolute cost of either direction.
Appendix~\ref{app:gas} reports the benchmark configuration and execution
environment.

\paragraph{Alternative protocol architectures.}
The same payoff equivalence can be operationalized at other protocol layers.
Hyperliquid exposes native outcome operations: \texttt{negateOutcome} realizes
the NO-to-YES direction, while \texttt{splitOutcome} and
\texttt{mergeQuestion} can be composed to realize the reverse direction
\cite{hyperliquidHIP4}. Starting with the supplied YES positions and
$(|S|-1)q$ collateral, the trader temporarily contributes an additional $q$,
splits $q$ collateral for every outcome in $S$, and merges the resulting
complete YES set. The merge restores the temporary $q$, leaving the
complementary NO basket. This construction has the same net collateral
requirement as Equation~\eqref{eq:reverse-realization}, but requires multiple
protocol operations and temporary working collateral rather than a dedicated
reverse primitive. Unless executed atomically, it may also expose the trader to
intermediate-state risk. Algorithm~\ref{alg:hyperliquid-reverse} formalizes the
composition.

Kalshi instead applies the payoff relation through account-level Collateral
Return. For eligible mutually exclusive and directional portfolios, it releases
collateral corresponding to payoff that has become redundant and adjusts the
remaining position value accordingly \cite{kalshiCollateralReturn}. This
reduces capital lock-up but does not transform YES positions and collateral into
a transferable complementary NO basket; collateral return may also restrict
subsequent sale of the affected positions.

These architectures expose payoff equivalence through, respectively, an atomic
token transformation, a composition of native operations, or account-level
netting. For pre-settlement enforcement, the relevant design dimensions are
therefore not only computational cost, but also atomicity, temporary collateral
requirements, outcome-set finality, and the transferability of the resulting
positions.

\section{Related Work}

\paragraph{Prediction-market efficiency and structured claims.}
Prediction-market efficiency is commonly studied as informational efficiency:
whether prices aggregate beliefs and forecast outcomes
\cite{wolfersZitzewitz2004,wolfersZitzewitz2006}. Hanson's work on
combinatorial information markets provides the foundation for trading logically
related claims \cite{hanson2003,hanson2007}. We study a complementary dimension:
whether prices satisfy deterministic payoff relations and whether protocols make
these relations enforceable before settlement. Our focus is therefore not only
payoff consistency, but protocol-executable no-arbitrage.

\paragraph{Polymarket microstructure and arbitrage.}
Recent work characterizes Polymarket's transaction structure and CLOB
microstructure \cite{tsangYang2026,dubach2026}. Cheng et al.\ reconstruct
executable order-book states and study same-market and cross-market
combinatorial arbitrage in NBA markets \cite{chengYangZou2026}. Whereas their
cross-market constraints arise from logical relations among NBA contracts, ours
are encoded by Polymarket's NegRisk market structure. Cheng et al.\ retain a
small number of same-market episodes as executable. Our independent WebSocket
validation shows that isolated deviations between complementary YES and NO
books can arise from API-level synchronization inconsistencies
(Appendix~\ref{app:websocket-validation}); we therefore exclude this channel
from our estimand.

Saguillo et al.\ estimate \$10.6 million from same-condition rebalancing and
\$29 million from rebalancing mutually exclusive NegRisk outcomes
\cite{saguilloEtAl2025}. Their opportunity analysis uses block-level VWAPs,
carrying prices forward for up to 5,000 blocks, while their realized-profit
reconstruction groups each address's fills within a 950-block window. For
NegRisk portfolios, they additionally permit low-probability outcomes to be
omitted and estimate the cost of the missing positions.

We instead target mechanism-linked realization. Converter-enabled arbitrage
requires an observed Adapter call whose inputs were acquired within five
blocks, while settlement-based arbitrage requires a complete basket formed
within ten minutes. We also exclude same-condition rebalancing because
Polymarket presents YES and NO as complementary views of a unified order book.
Restricting our analysis to the same 1 April 2024--1 April 2025 resolution
window yields \$291,424 under these criteria. The numerical difference therefore
reflects distinct estimands: broadly inferred rebalancing versus arbitrage
linked to an identifiable pre-settlement or settlement realization path.

\paragraph{Limits to arbitrage and realizability.}
More generally, the limits-to-arbitrage literature shows that payoff equivalence
need not produce immediate price convergence when realization is costly or
capital-constrained \cite{shleiferVishny1997}. Gebele et al.\ document the
cross-platform manifestation of this problem: semantically equivalent claims
cannot be netted across venues and exhibit persistent execution-aware price
differences \cite{semanticNonFungibility2026}. Related work shows that delayed
settlement generates maturity-dependent discounts and that NegRisk conversion
can reduce these discounts by enabling collateral recycling
\cite{settlementDiscount2026}. We hold the venue and event semantics fixed and
study the within-protocol mechanism directly: whether protocol-defined payoff
equivalences can be realized before settlement and how this affects their
enforcement.

\section{Conclusion}

Deterministic payoff equivalence does not, by itself, constitute an executable
no-arbitrage condition. How quickly and capital-efficiently it disciplines
prices depends on the realization mechanisms available before settlement.
Polymarket's negative-risk markets make this distinction observable: the
NegRisk Adapter operationalizes the NO-to-YES direction, whereas the reverse
direction remains settlement-dependent.

Consistent with this asymmetry, positive CLOB violations concentrate on the
unsupported YES side, while adapter-supported NO-side violations are
substantially less frequent and, in the limited duration sample, directionally
shorter-lived. At the actor level, we attribute 1.086 million USDC to
converter-enabled strategies, compared with 32,283 USDC from settlement-based
basket formation. Converter-enabled realization therefore accounts for 97\% of
estimated mechanism-linked profit, with most of this value arising from partial
conversions and subsequent inventory recycling. 

Our reverse-adapter prototype shows that the missing YES-to-NO path can be
implemented with approximately 63,000 additional gas units while preserving the
existing Adapter's linear scaling. Deployment nevertheless requires a finalized
active-outcome set, and absolute gas costs remain a constraint for events with
many outcomes. More broadly, payoff structure determines the economic
no-arbitrage bounds, while protocol architecture determines how effectively
traders can enforce them before settlement.

\clearpage

\bibliographystyle{splncs04}
\bibliography{references}

\clearpage
\appendix

\section{Same-Condition WebSocket Validation}
\label{app:websocket-validation}

To assess whether same-condition YES/NO deviations represent durable arbitrage
or synchronization artifacts, we replicated the Polymarket market WebSocket
over the seven-day window 8 May 2026 15:00 UTC--15 May 2026 15:00 UTC. From
308,416,666 messages, we reconstructed synchronized YES and NO order books and
compared every entry with its complementary entry on the opposite book side.
The two books were exact mirrors except for 20 isolated API-level
inconsistencies. In each of these cases, only one complementary pair failed to
match; all other entries in the same books remained consistent. This provides
strong evidence that residual same-condition deviations reflect feed or API
synchronization artifacts rather than a durable executable arbitrage channel.

\section{Conversion Algorithms}
\label{app:algorithms}

\begin{algorithm}[H]
    \small
    \caption{Existing NegRisk Adapter (\texttt{convertPositions}): NO-to-YES}
    \label{alg:no-to-yes}
    \begin{algorithmic}[1]
        \Require Binary-market set $Q$ of a negative-risk event
        \Require Caller-supplied $q>0$ units of each $\mathrm{NO}_i$
        for a nonempty input set $S\subseteq Q$

        \State $C\gets Q\setminus S$
        \Comment{YES-output markets}
        \State $q_{\mathrm{fee}}\gets
        \left\lfloor q\cdot\texttt{feeBips}/10{,}000\right\rfloor$
        \Comment{fee per output leg}
        \State $q_{\mathrm{out}}\gets q-q_{\mathrm{fee}}$
        \Comment{net quantity per output leg}

        \State the Adapter mints $|C|q$ units of wrapped collateral
        \Comment{created internally}
        \ForAll{$j\in C$}
            \State split $q$ units of wrapped collateral into
            $q$ units each of $(\mathrm{YES}_j,\mathrm{NO}_j)$
        \EndFor

        \ForAll{$i\in S$}
            \Comment{caller-supplied NO positions}
            \State transfer $q$ units of $\mathrm{NO}_i$
            from the caller to the non-redeemable burn address
        \EndFor
        \ForAll{$j\in C$}
            \Comment{NO positions created above}
            \State transfer $q$ units of $\mathrm{NO}_j$
            from the Adapter to the non-redeemable burn address
        \EndFor

        \State transfer $(|S|-1)q_{\mathrm{out}}$ collateral to the caller
        \State transfer $(|S|-1)q_{\mathrm{fee}}$ collateral to the fee vault

        \ForAll{$j\in C$}
            \Comment{YES positions created above}
            \State transfer $q_{\mathrm{out}}$ units of $\mathrm{YES}_j$
            from the Adapter to the caller
            \State transfer $q_{\mathrm{fee}}$ units of $\mathrm{YES}_j$
            from the Adapter to the fee vault
        \EndFor
    \end{algorithmic}
\end{algorithm}

Algorithms~\ref{alg:no-to-yes} and~\ref{alg:yes-to-no} summarize the two
conversion directions. The existing Adapter converts caller-supplied NO
positions on $S$ into $(|S|-1)q$ collateral and YES positions on
$C=Q\setminus S$ (Algorithm~\ref{alg:no-to-yes}). The proposed reverse path
converts YES positions on $Y$, together with $(|N|-1)q$ collateral when
$N=Q\setminus Y$ is nonempty, into NO positions on $N$; a complete YES set
instead returns collateral (Algorithm~\ref{alg:yes-to-no}). Both paths rely on an Adapter-controlled collateral wrapper. The Adapter
synthetically mints wrapped collateral for the required YES/NO pairs and makes
the consumed input positions and internally created non-output positions
non-redeemable. Collateral conservation follows from the corresponding payoff
identities, and the implementation depends on the Adapter's ability to mint
wrapped collateral and release the underlying collateral held by the wrapper.

\begin{algorithm}[H]
    \small
    \caption{Proposed Reverse Adapter (\texttt{convertYESPositions}): YES-to-NO}
    \label{alg:yes-to-no}
    \begin{algorithmic}[1]
        \Require Binary-market set $Q$ of a negative-risk event
        \Require Caller-supplied $q>0$ units of each $\mathrm{YES}_j$
        for a nonempty input set $Y\subseteq Q$

        \State $N\gets Q\setminus Y$
        \Comment{NO-output markets}
        \State $q_{\mathrm{fee}}\gets
        \left\lfloor q\cdot\texttt{feeBips}/10{,}000\right\rfloor$
        \Comment{fee per output leg}
        \State $q_{\mathrm{out}}\gets q-q_{\mathrm{fee}}$
        \Comment{net quantity per output leg}

        \If{$N=\varnothing$}
            \Comment{complete YES input}
            \ForAll{$j\in Y$}
                \State transfer $q$ units of $\mathrm{YES}_j$
                from the caller to the non-redeemable burn address
            \EndFor
            \State transfer $q_{\mathrm{out}}$ collateral to the caller
            \State transfer $q_{\mathrm{fee}}$ collateral to the fee vault
            \State \Return
        \EndIf

        \State the Adapter mints $|N|q$ units of wrapped collateral
        \Comment{created internally}
        \ForAll{$i\in N$}
            \State split $q$ units of wrapped collateral into
            $q$ units each of $(\mathrm{YES}_i,\mathrm{NO}_i)$
        \EndFor

        \ForAll{$j\in Y$}
            \Comment{caller-supplied YES positions}
            \State transfer $q$ units of $\mathrm{YES}_j$
            from the caller to the non-redeemable burn address
        \EndFor
        \ForAll{$i\in N$}
            \Comment{YES positions created above}
            \State transfer $q$ units of $\mathrm{YES}_i$
            from the Adapter to the non-redeemable burn address
        \EndFor

        \State collect and wrap $(|N|-1)q$ collateral from the caller

        \ForAll{$i\in N$}
            \Comment{NO positions created above}
            \State transfer $q_{\mathrm{out}}$ units of $\mathrm{NO}_i$
            from the Adapter to the caller
            \State transfer $q_{\mathrm{fee}}$ units of $\mathrm{NO}_i$
            from the Adapter to the fee vault
        \EndFor
    \end{algorithmic}
\end{algorithm}

\subsection{Native Composition on Hyperliquid}
\label{app:hyperliquid-conversion}

Hyperliquid does not expose a direct YES-to-NO operation, but the reverse
realization path can be composed from its native outcome operations. The
composition temporarily requires $q$ additional units of quote collateral,
which are restored when the resulting complete YES set is merged. Its net
collateral consumption is therefore $(|N|-1)q$.

\begin{algorithm}[t]
\caption{Hyperliquid Composition for YES-to-NO Realization}
\label{alg:hyperliquid-reverse}
\begin{algorithmic}[1]
\Require Active outcome set $Q$
\Require Caller-supplied $q>0$ units of each $\mathrm{YES}_j$ for a nonempty
         input set $Y\subseteq Q$
\State $N \gets Q\setminus Y$
\If{$N=\emptyset$}
    \State Call $\texttt{mergeQuestion}(Q,q)$ using the complete YES set
    \State \Return $q$ units of quote collateral
\EndIf
\State Add $(|N|-1)q$ units of quote collateral
       \Comment{required net top-up}
\State Temporarily add $q$ units of quote collateral
       \Comment{working collateral}
\ForAll{$i\in N$}
    \State Call $\texttt{splitOutcome}(i,q)$
    \Comment{receive $q\,\mathrm{YES}_i$ and $q\,\mathrm{NO}_i$}
\EndFor
\State Call $\texttt{mergeQuestion}(Q,q)$ using all supplied and created YES shares
\State Withdraw the returned $q$ quote units
       \Comment{restore working collateral}
\State \Return $q$ units of each $\mathrm{NO}_i$ for $i\in N$
\end{algorithmic}
\end{algorithm}

\section{Conversion Gas Benchmarks}
\label{app:gas}

\begin{table}[H]
\centering
\small
\begin{tabular}{r@{\hspace{1.5em}}r@{\hspace{2em}}r@{\hspace{2em}}r}
\toprule
\textbf{Outcomes} &
\multicolumn{1}{c}{\textbf{NO-to-YES}} &
\multicolumn{1}{c}{\textbf{YES-to-NO}} &
\multicolumn{1}{c}{\textbf{Additional}} \\
&
\multicolumn{1}{c}{\texttt{convertPositions}} &
\multicolumn{1}{c}{\texttt{convertYESPositions}} &
\multicolumn{1}{c}{\textbf{gas}} \\
\midrule
5   &    638,077 &    701,171 & 63,094 \\
32  &  3,943,631 &  4,006,726 & 63,095 \\
64  &  8,172,226 &  8,235,320 & 63,094 \\
96  & 12,215,519 & 12,278,614 & 63,095 \\
128 & 16,438,894 & 16,501,989 & 63,095 \\
196 & 25,253,599 & 25,316,694 & 63,095 \\
\bottomrule
\end{tabular}
\caption{Gas usage for existing NO-to-YES and proposed YES-to-NO conversion paths.}
\label{tab:bidir-gas}
\end{table}

\section{Conversion Diagnostics}
\label{app:conversion-diagnostics}

\paragraph{Anomalous conversion cluster exclusion.}
The main realized-profit analysis excludes a single address-date cluster of
conversion transactions. The exclusion is restricted to 75 conversions executed
between 22:07 and 22:14 UTC on 2025-12-12 by three stakeholder addresses:
\href{https://polygonscan.com/address/0x8E9Eedf20DfA70956d49F608a205e402d9df38e4}{0x8E9E...38e4},
\href{https://polygonscan.com/address/0x5F390E4B7D6f06D6756a6c92Afdbf7b3176aa78c}{0x5F39...a78c},
and
\href{https://polygonscan.com/address/0xcd91A549956854FDc38efad7e80fF5da8F5432b8}{0xcd91...32b8}.

These transactions are structurally distinct from the remainder of the
high-profit conversion sample. They exhibit unusually low capital deployment,
extreme return-on-cost ratios, and realized profits that are almost entirely
explained by the core conversion component. In addition, midpoint prices before
and after the conversions remain close to 50 cents, suggesting that thin or
empty order books were temporarily populated with artificial orders and then
consumed to create mechanically profitable conversion sequences.

We therefore interpret this cluster as non-representative conversion activity,
plausibly reflecting value transfer between related addresses rather than
competitive arbitrage. The exclusion affects only this address-date cluster and
does not materially affect the qualitative conclusions of the paper. All main
results are robust to replacing the address-date filter with purely diagnostic
thresholds based on extreme return-on-cost ratios and atypical core-to-total
profit decompositions.

\section{Weekly Maker--Taker Composition}
\label{app:weekly-maker-taker-composition}

\begin{figure}[t]
\centering
\includegraphics[width=\textwidth]{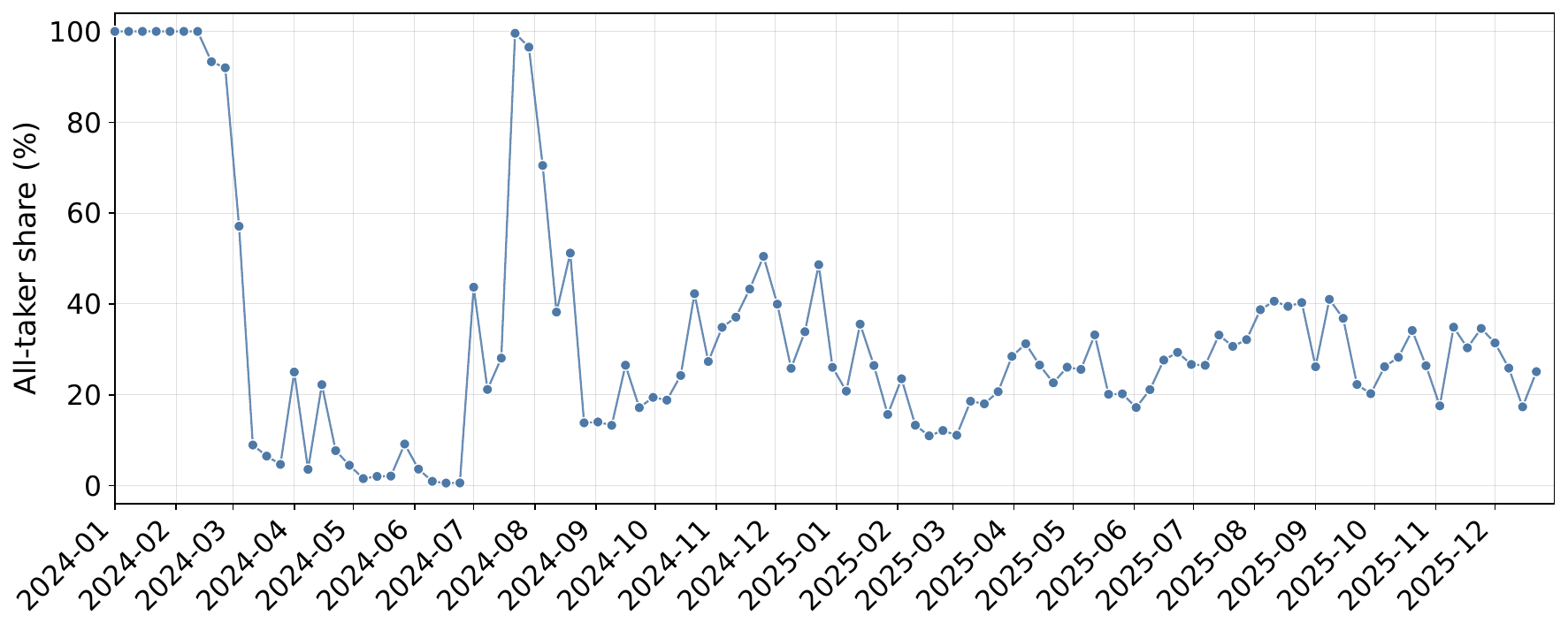}
\caption{Weekly share of analyzed conversions executed entirely through taker
fills. The complementary share consists of maker-assisted conversions.}
\label{fig:weekly-maker-taker-share}
\end{figure}

Figure~\ref{fig:weekly-maker-taker-share} reports the weekly execution mix of
the analyzed conversions. The decline in the all-taker share indicates the
increasing prevalence of maker-assisted conversion strategies.

\section{Weekly Converter Profit per Trade}
\label{app:weekly-profit-per-trade}

\begin{figure}[t]
\centering
\includegraphics[width=\textwidth]{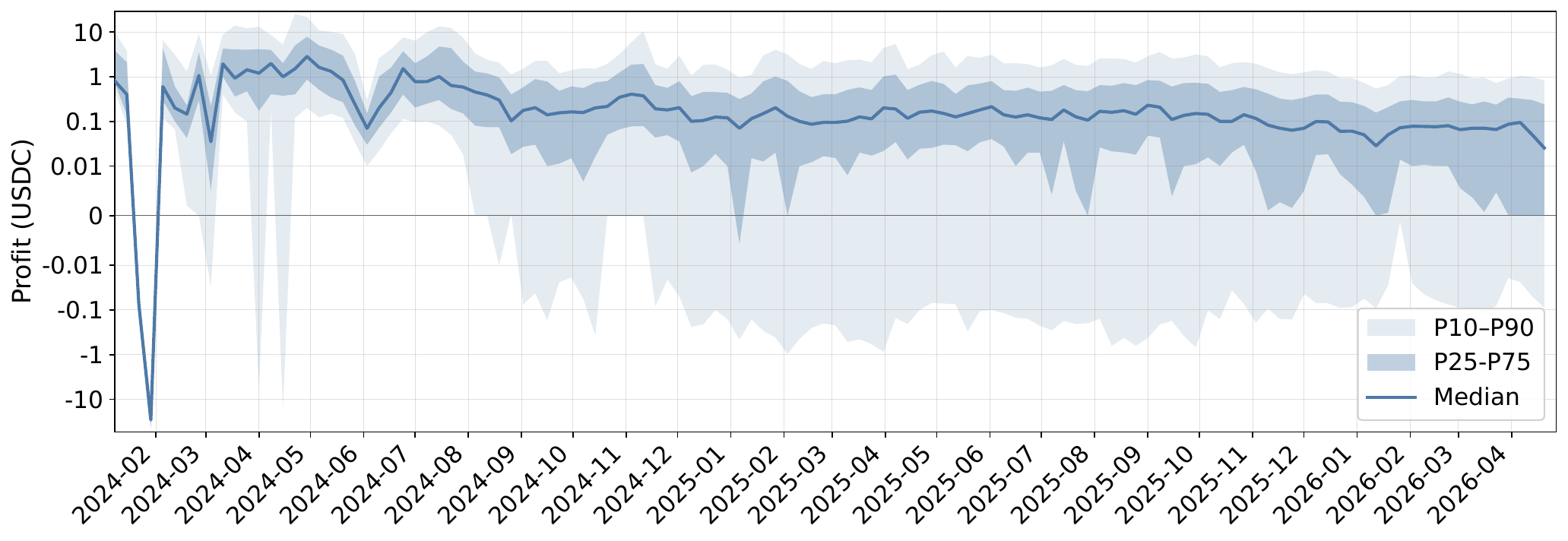}
\caption{Weekly distribution of realized converter-enabled profit per trade.
The line shows the median, and the shaded bands show the 25th--75th and
10th--90th percentiles.}
\label{fig:weekly-profit-per-trade-percentiles}
\end{figure}

Figure~\ref{fig:weekly-profit-per-trade-percentiles} shows how the distribution
of realized converter-enabled profit per trade evolves over time.

\section{Settlement-Based Arbitrage Over Time}
\label{app:settlement-basket-distribution}

\begin{figure}[t]
\centering
\includegraphics[width=\textwidth]{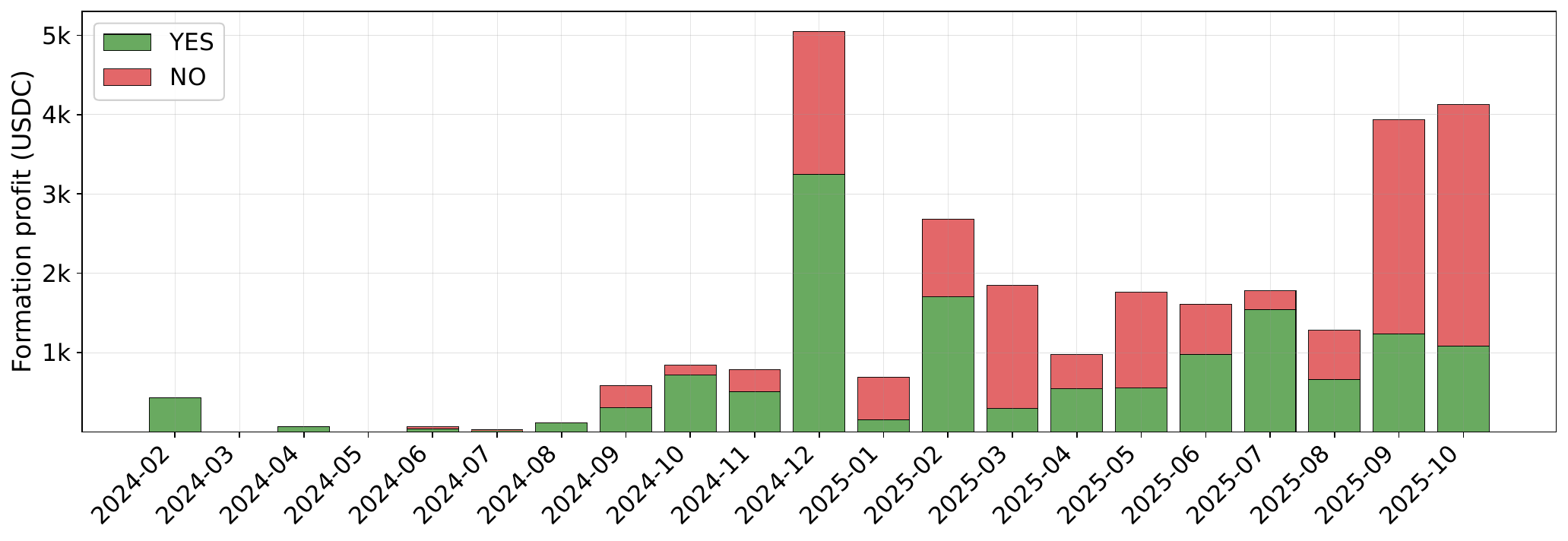}
\caption{Monthly settlement-based arbitrage profit by complete-basket formation
month, February 2024 to October 2025.}
\label{fig:settlement-monthly-formation-profit}
\end{figure}

Figure~\ref{fig:settlement-monthly-formation-profit} complements the aggregate
settlement-based results by showing when complete baskets were formed.

\end{document}